\documentclass[a4paper,UKenglish,cleveref,autoref,thm-restate]{lipics-v2021}

\EventEditors{Robbert Krebbers and Alexandra Silva}
\EventNoEds{2}
\EventLongTitle{40th European Conference on Object-Oriented Programming (ECOOP 2026)}
\EventShortTitle{ECOOP 2026}
\EventAcronym{ECOOP}
\EventYear{2026}
\EventDate{June 29--July 3, 2026}
\EventLocation{Brussels, Belgium}
\EventLogo{}
\SeriesVolume{372}
\ArticleNo{16}

\hypersetup{colorlinks=true, linkcolor=blue, citecolor=blue, urlcolor=blue}

\usepackage{algorithmic}
\usepackage[linesnumbered,ruled,vlined]{algorithm2e}
\usepackage{textcomp}
\usepackage{booktabs}
\usepackage{multirow, multicol}
\usepackage{makecell}
\usepackage{tikz}
\usetikzlibrary{shapes.geometric, positioning}
\usepackage{pifont} 
\usepackage{threeparttable}
\usepackage{adjustbox}
\usepackage{pgfplots}
\usepackage{balance} 
\usepackage{listings}
\usepackage{resource/cstyle}
\pgfplotsset{compat=1.18} 

\usepackage{subcaption}

\usepackage{comment}

\definecolor{darkgreen}{rgb}{0, 0.5, 0}
\definecolor{darkred}{rgb}{0.6, 0, 0}
\definecolor{lightblue}{rgb}{0.7, 0.7, 0.9}
\definecolor{lightred}{rgb}{1, 0.75, 0.75}
\newcommand{\cmark}{\ding{51}}
\newcommand{\xmark}{\ding{55}}

\newcommand{\vNO}{\textcolor{gray}{\xmark}}

\newcommand{\vYES}{\textcolor{darkgreen}{\cmark}}
\newcommand{\vvYES}{\textcolor{darkred}{\cmark}}

\newcommand{\blue}[1]{{#1}}

\newcommand{\binsec}{BINSEC\xspace}

\newcommand{\jingbo}[1]{}
\renewcommand{\jingbo}[1]{{\color{teal}{\bf Jingbo says:#1}}}

\title{Efficient Symbolic Execution of Software under Fault Attacks}

\author{Yuzhou Fang}{University of Southern California, Los Angeles, USA}{yuzhoufa@usc.edu}{https://orcid.org/0000-0002-4933-7443}{}

\author{Chenyu Zhou}{University of Southern California, Los Angeles, USA}{czhou691@usc.edu}{https://orcid.org/0009-0006-8493-6886}{}

\author{Jingbo Wang}{Purdue University, West Lafayette, USA}{wang6203@purdue.edu}{https://orcid.org/0000-0001-5877-2677}{}

\author{Chao Wang}{University of Southern California, Los Angeles, USA}{wang626@usc.edu}{https://orcid.org/0009-0003-4684-3943}{}

\authorrunning{Y. Fang, C. Zhou, J. Wang, and C. Wang} 

\Copyright{Yuzhou Fang, Chenyu Zhou, Jingbo Wang, and Chao Wang} 



\funding{This work was supported in part by the NSF grant CCF-2220345. }

\nolinenumbers 

\begin{document}

\maketitle

\begin{abstract}
We propose a symbolic execution method for analyzing the safety of software under fault attacks both accurately and efficiently. 
Fault attacks leverage physically injected hardware faults in an embedded system to break the safety of a software program. 
While there are existing methods for analyzing the impact of maliciously injected hardware faults on the embedded software, they suffer from inaccurate fault modeling and inefficient fault analysis.  To overcome these limitations, we propose two novel techniques. 
First, we propose a new fault modeling technique that leverages automated program transformation to add symbolic variables to the original program, to accurately model the new program behavior induced by the injected faults.  
This new fault modeling approach has two advantages over existing techniques: (a) the fault-induced program behavior is closely related to what attackers exploit in practice and (b) the automatically transformed program may be analyzed by any downstream fault analysis algorithm.
Second, we propose an efficient symbolic execution algorithm that is designed specifically for conducting fault analysis on the transformed program.  It leverages two pruning techniques to mitigate path explosion, which is the main performance bottleneck of symbolic execution in general and, in this particular application, is exacerbated by the additional fault-induced program behavior. 
We have implemented the proposed method and evaluated it on a variety of benchmark programs. 
The experimental results show that our method significantly outperforms the state-of-the-art techniques. Specifically, our method not only drastically reduces the overall running time of symbolic execution but also retains its error detection capabilities.  Compared to the current state-of-the-art, it is able to detect previously-missed safety violations and at the same time avoid bogus violations. 
Furthermore, compared to the baseline algorithm, our optimized symbolic execution algorithm can be orders-of-magnitude faster. 
\end{abstract}

\section{Introduction}
\label{sec:intro}

Fault injection is a technique that attackers often use in the context of embedded systems to violate the safety property of a software program by injecting hardware faults into the underlying CPU hardware. In practice, this has been accomplished using various physical mechanisms including clock glitching~\cite{ClockGlitch18}, voltage glitching~\cite{PowerGlitching05}, electromagnetic (EM) waves~\cite{OFlynn20}, and laser beams~\cite{LaserBeam03}.
Prior studies~\cite{Bar-ElCNTW06,YuceSW18,GrycelS21,LiuSS24} have shown that such maliciously injected hardware faults can lead to abnormal behaviors in a software program that runs on top of the hardware. Furthermore, these abnormal program behaviors are often statistically predictable and, in some cases, are even somewhat controllable by attackers, e.g., in the form of targeted instruction skipping, where an \textsf{if}-statement that guards access to some sensitive data is skipped, thus leading to safety violations.

Conventional techniques for assessing the potential impact of such hardware fault injection attacks on the embedded software rely primarily on simulation techniques.  Broadly speaking, these techniques include hardware simulation, software simulation, and hardware-software co-simulation. However, these simulation-based techniques tend to be slow. For example, hardware simulation requires the availability of a detailed model of the CPU, and the modeling of fault injection/propagation, often at the register-transfer level (RTL).  RTL simulation is extremely slow. Furthermore, all simulation techniques suffer from limited behavior coverage, because they run the software program on the hardware model for one specific program input at a time.
In contrast, symbolic execution-based techniques \emph{have the potential} to be more efficient by simultaneously covering all program inputs and execution paths, including all of the fault-induced execution paths, provided that the fault impact is modeled accurately. However, existing symbolic execution-based techniques for fault analysis tend to suffer from inaccurate fault modeling and inefficient fault analysis.

\vspace{1.5ex}\noindent
\textbf{Fault Modeling.}
Current state-of-the-art techniques for modeling the impact of hardware faults on a software program have two limitations. First, the fault model is not realistic in the sense that it does not model what actually happens at the hardware level. Second, the fault model is not accurate in the sense that it may miss fault-induced program behaviors that can show up in practice. 
For example, to model the impact of a hardware fault on the program's control flow, an existing technique is called \emph{test inversion}~\cite{ESOP23}. 
In general, a program may implement a branching statement, such as an \textsf{if-else} statement, using a pair of comparison and jump operations.  Specifically, the comparison sets a CPU flag and the jump operation tests the CPU flag, based on which the control is transferred to different program locations 
(see Section~\ref{subsec:testinversion} for details).
Test inversion assumes that, right before testing, the CPU flag may be inverted from $true$ to $false$ or vice versa, depending on whether a hardware fault is injected at that moment. Thus, the impact of the hardware fault is modeled by flipping the two conditions guarding the two branches of an \textsf{if-else} statement. 

While this may appear to be a reasonable approach at first glance, since it does capture some of the fault-induced execution paths, the model does not reflect what actually happens inside the CPU hardware.  For example, it would be extremely difficult in practice for an attacker to flip a CPU flag in a predictable and targeted manner. Therefore, it is not how a real attack is carried out.  
Instead, what typically happens in practice is that a jump instruction associated with the \textsf{if-else} statement is turned into a \textit{nop} (e.g., due to bit-flips in the jump instruction's encoding, which often occur during faulty instruction fetching).  As a result, the jump instruction is skipped.
We will show later in this paper (both in Section~\ref{sec:modeling} and during the experimental evaluation in Section~\ref{sec:experiment}) that instruction skipping is significantly more accurate than test inversion: it can detect realistic safety violations that otherwise would have been overlooked.

\vspace{1.5ex}\noindent
\textbf{Fault Analysis.}
Another limitation of the current state-of-the-art techniques is the inefficiency of their fault analysis algorithms.  Since the impact of a hardware fault may show up at any moment during the execution of a software program, all possible fault-induced execution paths must be analyzed to check if any of them may lead to the violation of a safety or security property. 
While symbolic execution is capable of exploring all program paths for all program inputs, including the fault-induced execution paths that otherwise would not have been feasible, it is also well known that symbolic execution suffers from the \emph{path explosion} problem. When hardware faults are injected, the path explosion problem is further exacerbated.  
While there are existing techniques for mitigating the path explosion problem, they are not designed specifically for fault analysis.

The transformed program subjected to a fault analysis has some unique characteristics.  First, the symbolic variables that we introduce for modeling the impact of hardware faults are significantly different from ordinary program variables; treating them in the same manner during symbolic execution can make it difficult for pruning algorithms to identify redundant execution paths.  Furthermore, there can be subtle interactions between the two types of variables.  
Furthermore, existing methods often impose a bound on the maximum number of faults that an execution can activate using a technique called \emph{fault saturation}~\cite{ESOP23}, which avoids exploring paths where the number of faults exceeds the bound.  While this technique can reduce the number of explored paths, by itself, it is not sufficient for taming the often astronomically large number of execution paths.  Instead, its tight integration with techniques for pruning redundant paths must be considered.

\vspace{1.5ex}\noindent
\textbf{Proposed Method.}
To overcome the two limitations mentioned above, we propose a new symbolic execution method to analyze the impact of hardware faults on a software program both accurately and efficiently. 
Our method has two innovations. 
The first one is a program transformation technique for accurately modeling the fault-induced program behavior. 
The second one is a symbolic execution algorithm for efficient fault analysis, equipped with a combination of fault saturation and redundancy pruning.

Unlike existing methods where fault modeling is part of the fault analysis procedure, our method implements fault modeling as a standalone program transformation. It leverages a compiler to automatically insert auxiliary variables into the program at the intermediate representation (IR) level, to model the impact of faults on the program's control flow.
Our fault modeling technique is realistic in the sense that it faithfully implements the observations reported by prior empirical studies~\cite{YuceSW18,GrycelS21,LiuSS24}, which show that the primary source of fault attacks in practice is due to the fact that hardware faults can lead to jump instructions being skipped. 
Thus, for each jump instruction in the original program, we add some additional code to create an auxiliary variable at run time, called the \emph{fault flag}. 
%
%
Details will be presented in Section~\ref{sec:modeling}. For now, it suffices to understand that, if the jump instruction is inside a loop,
each time the jump instruction is executed, a separate \emph{fault flag} is created.  
When all of the \emph{fault flags} are disabled, for example, the transformed program behaves exactly the same as the original program. 
However, when some \emph{fault flags} are enabled, the transformed program may have additional behaviors; they correspond to some jump instructions being skipped during program execution.

Our second innovation is an efficient symbolic execution algorithm for conducting fault analysis. Its efficiency comes from a new technique for mitigating the path explosion problem. 
The technique combines a path summary-based analysis with fault saturation to identify and then avoid redundant execution paths. 
In this context, our main observation is that multiple execution paths often share the same suffix. While such sharing is common among fault-free execution paths, it is even more common among fault-induced execution paths, as well as between fault-free execution paths and fault-induced execution paths. 
In such cases, the shared suffix may only need to be symbolically executed once, instead of being symbolically executed again along different paths (prefixes).

Specifically, the weakest precondition (WP~\cite{dijkstra1976wp}) computed along a previously-explored suffix can serve as a sound and effective summary of the partial path (the suffix).  We leverage this path summary to define a sufficient condition, under which the symbolic execution of this shared suffix will not lead to any previously-unseen violations.  
Thus, when the shared suffix is encountered again, e.g., following the symbolic execution of another program path (another prefix), we can safely skip the symbolic execution of the shared suffix, because it is deemed redundant. 
In other words, skipping the partial path would not negatively affect the symbolic execution algorithm's ability to detect safety violations.

\vspace{1.5ex}\noindent
\textbf{Evaluation.}
To find out whether the proposed method is effective in practice, we have implemented the method in a software tool by leveraging three existing tools:  the LLVM compiler~\cite{LLVM04}, the KLEE symbolic execution engine~\cite{KLEE08}, and the Z3 SMT solver~\cite{Z308}.  
The software tool is designed to analyze C programs where the safety property is specified using an assertion, which is expected to hold for all program paths and program inputs.  Alternatively, the safety property may be regarded as a program location named ERROR, which should never be reached.
%
Given such a C program as input, our tool leverages the LLVM compiler's existing capability to compile C code to LLVM bitcode, and then applies our fault modeling to LLVM bitcode.  Next, it leverages our extension of KLEE to
conduct efficient symbolic execution, to detect violations of the safety property represented as an assertion embedded in the program.  The tool also leverages Z3 to perform the symbolic analysis needed for identifying and pruning the redundant execution paths.

For experimental evaluation, we have implemented both our method and a state-of-the-art method from~\cite{ESOP23} in the same tool.  We have applied both methods to 112 benchmark programs.  These benchmark programs come from two sources.  The first source is the VerifyPIN benchmark suite used by the existing method~\cite{ESOP23}. The second source is the \emph{array-crafted}, \emph{bitvector}, and \emph{bitvector-loops} categories of the SV-COMP suite~\cite{SV-COMP24}.
We have also conducted the ablation study by applying our method with and without the new redundancy pruning technique.  
The experimental results show that our method significantly outperforms the current state-of-the-art. Specifically, our method can not only drastically reduce the running time taken by fault analysis, but also can detect previously-missed safety violations. Compared to the baseline symbolic execution algorithm, our new symbolic execution method, augmented with redundancy pruning, can be orders of magnitude faster.

\vspace{1.5ex}
%
To summarize, this paper makes the following technical contributions: 
\begin{itemize}
\item 
\blue{We propose a symbolic execution method for accurately and efficiently analyzing the safety property of a software program running on a hardware platform that is subjected to fault attacks.} 
\item
\blue{We design an automated program transformation to add auxiliary symbolic variables and encode fault-induced control flows, to allow the transformed program to model fault-induced program behaviors.}
\item
\blue{We design a redundancy pruning technique, which leverages both fault saturation and a weakest precondition based path summary to identify and eliminate more redundant paths during symbolic execution.}
\item 
\blue{We implement the method and demonstrate its superiority over the current state-of-the-art on two sets of widely used benchmark programs.}
\end{itemize}

The remainder of this paper is organized as follows. We provide the technical background in Section~\ref{sec:background}. Then, we present the top-level procedure of our symbolic execution method in Section~\ref{sec:methodology}. This is followed by our fault modeling technique in Section~\ref{sec:modeling} and our new redundancy pruning technique in Section~\ref{sec:pruning}. After that, we present the experimental results in Section~\ref{sec:experiment} and review the related work in Section~\ref{sec:related}. Finally, we give our conclusion in Section~\ref{sec:conclusion}.

\section{Background}
\label{sec:background}

We first use an example software program
to illustrate the limitations of existing methods,
and then
present the threat model of hardware fault attacks manifested on embedded software.

\subsection{The Example Program}
\label{subsec:motivation}

\begin{figure*}[t]
    \centering
    \includegraphics[width=\textwidth]{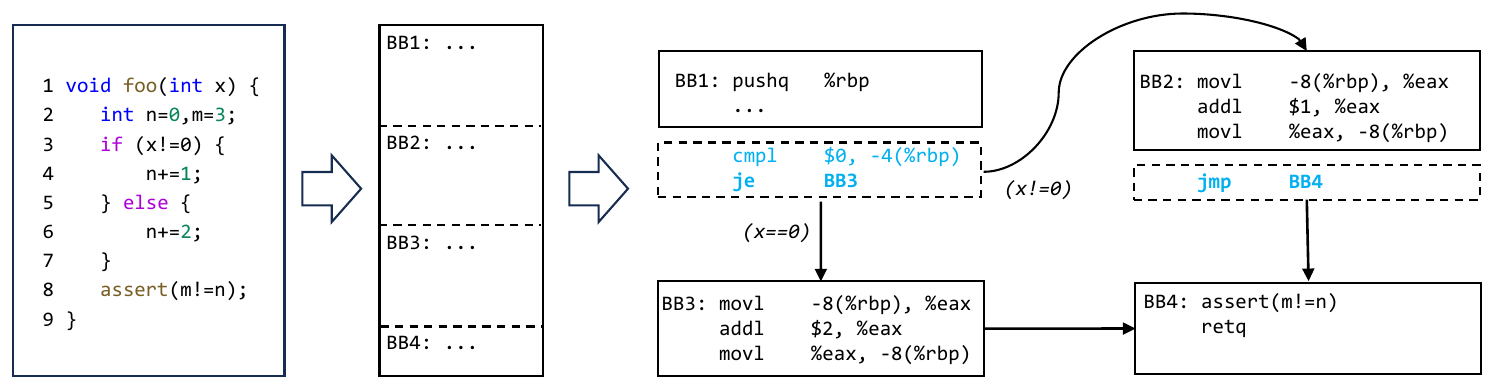}
    \caption{An example C program (left) and its x86 assembly code shown as a sequence of basic blocks (middle) and the CFG (right). Note that the assertion at Line~8 always holds.}
    \label{fig:motivating}
\end{figure*}

Fig.~\ref{fig:motivating} shows a victim software program, where the C code is on the left-hand side.  Depending on the value of input variable \texttt{x}, the function \texttt{foo(x)} may execute either of the two branches in the \textsf{if-else} statement. For all program paths and program inputs, the assertion at Line~8 always holds. 
In the middle are the four basic blocks shown in the program's x86-64 assembly code, produced by the Clang/LLVM compiler. They are also the same basic blocks \texttt{BB1-BB4} shown in the control flow graph (CFG) on the right-hand side of Fig.~\ref{fig:motivating}.

Specifically, the conditional jump instruction (\texttt{je BB3}) corresponds to \texttt{if(x!=0)} in the C program. Depending on the value of \texttt{x}, the CPU will execute either \texttt{BB2} or \texttt{BB3}. 
The unconditional jump instruction (\texttt{jmp BB4}) redirects the control from \texttt{BB2} to \texttt{BB4}.

\subsubsection{Fault Modeling}
\label{subsec:testinversion}
%
The example program in Fig.~\ref{fig:motivating} illustrates the inaccuracy of fault modeling in the current state-of-the-art~\cite{ESOP23}, which relies on \emph{test inversion}.  That is, the fault is modeled by inverting the CPU flag used by the conditional jump instruction.
When the fault is injected at the end of \texttt{BB1}, in particular, the resulting CPU flag set by the comparison instruction \texttt{cmpl \$0,-4(\%rbp)} will be inverted, which has the effect of flipping the \textsf{if-else} branches in the C program. 
However, this does not lead to the violation of the assertion in Line~8, because as we have mentioned earlier, the assertion always holds, regardless of which of the two branches is executed. 

In contrast, our fault modeling allows the assertion in Line~8 to be violated, and thus matches what may actually happen in practice (the ground truth).  
This is because our technique relies on \emph{branch instruction skipping} to model the fault.  When the fault is injected at the end of \texttt{BB2}, it turns the jump instruction \texttt{jmp BB4} into a null operation (\textit{nop}).
Consequently, the program enters \texttt{BB3} right after \texttt{BB2}.  This has the effect of executing \emph{both branches} of the \textsf{if-else} statement.
As a result of our fault modeling, the value of $n$ becomes 3 at Line~8, which violates the assertion condition \texttt{(m!=n)}.
 
Note that the violation would not have been detected by \emph{test inversion}, since it relies on inverting the CPU flag, but the unconditional jump instruction \texttt{jmp BB4} does not check the CPU flag.

\subsubsection{Path Pruning}
%
The example program in Fig.~\ref{fig:motivating} also illustrates the inefficiency of existing fault analysis algorithms.  
For the control flow graph (CFG) on the right-hand side of Fig.~\ref{fig:motivating}, there will be five distinct execution paths using our fault modeling with transformation (the five paths will be shown in Fig.~\ref{fig:motivating_transformed}).   Thus, a baseline symbolic execution algorithm would explore all five paths.
Some existing methods rely on \emph{fault saturation}~\cite{ESOP23} to reduce the number of explored paths, but the technique is not always effective. In fact, it is largely ineffective for the example program shown in Fig.~\ref{fig:motivating}:
even with \emph{fault saturation}, symbolic execution would still completely execute four paths if one fault is allowed, while only one path (Path\#5) that requires two faults is partially pruned.

In contrast, our symbolic execution algorithm will explore only two complete paths (they are the first two paths shown in Fig.~\ref{fig:motivating_transformed}). One path has an assertion violation, and the other path does not.
The remaining three paths are only partially explored before our method identifies them as \emph{redundant} with respect to the already explored paths, and thus skips the remaining paths (suffixes).

While in this small example program, a reduction from five paths to two paths (together with three partially explored paths) may seem modest, the reduction will become more significant on larger programs. 
This is because the number of paths in a program tends to grow exponentially as the size of the program increases. 
During the experimental evaluation in Section~\ref{sec:experiment}, we will show that, in practice, the reduction can be as high as several orders of magnitude.

\subsection{The Threat Model}
\label{subsec:faultmodel}

Regarding hardware fault attacks manifested on embedded software, 
the threat model includes the goal and the capability of the adversary, as well as the fault-induced behavior of the victim software program. 

\subsubsection{The Victim Software Program}

Let $P$ be the victim software program and $\phi$ be a safety property of program $P$, represented as an assertion statement embedded in $P$.   Furthermore, during a fault-free execution of the program $P$, the property $\phi$ is expected to hold for all program paths and for all program inputs.

\subsubsection{The Goal and Capability of the Adversary}

The goal of the adversary is to violate the property $\phi$ by physically injecting faults into the underlying CPU hardware, thus changing the victim program's behavior. Specifically, the goal is to violate the safety property $\phi$.
%
While faults may be injected in various ways according to the literature~\cite{PowerGlitching05,ClockGlitch18,LaserBeam03}, due to physical limitations, there is often a bound ($\beta$) on the maximum number of faults that can be injected during each execution of the victim program. 
In this paper, the bound $\beta$ is called the \emph{fault budget}. 

\subsubsection{Fault-Induced Program Behavior}

Although maliciously injected hardware faults may lead to various abnormal behaviors in the victim software program, not all of them are predictable or controllable.  If they are not predictable and controllable, they will not be easily exploited and thus will not be practically useful to an adversary.
%
Furthermore, if the abnormal behavior is a moderate deviation, it is generally more useful than a drastic deviation. For example, if the deviation is so drastic that it even leads to the crash of the victim program (or the crash of the entire computer), it may become ineffective for the purpose of stealing information or secretly gaining control. 

With this in mind, prior studies of hardware fault-based security attacks~\cite{YuceSW18,GrycelS21,LiuSS24} tend to focus more on fault-induced  \emph{branch instruction skipping}.  In contrast, they do not focus on arbitrary data corruption since such corruption is less predictable or controllable.
It is worth noting that (silent) data corruption is a significant problem in large-scale infrastructure systems, high-performance computing applications, and cloud data centers.  However, it is out of the scope of this paper.

\subsubsection{Branch Instruction Skipping}

\begin{figure*}[t]
    \centering
    \includegraphics[width=\textwidth]{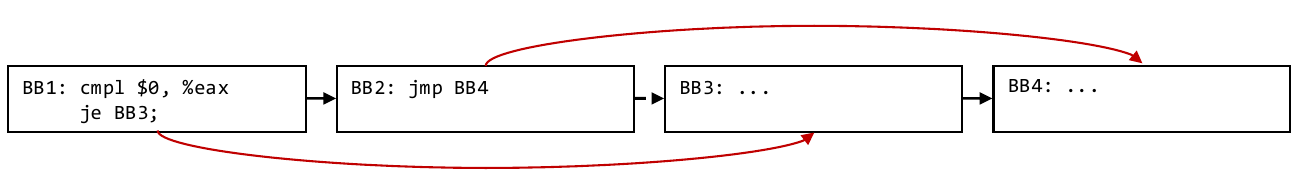}
    \caption{The control flow of an x86 instruction sequence, where the red arrows indicate the control flow brought by the jump instructions and the dashed arrows indicate the possibly sequential execution under a fault.}
    \label{fig:x86_BB}
\end{figure*}

Due to maliciously injected faults, the CPU hardware may interpret a jump instruction (\texttt{jmp}) as a null operation (\textit{nop}).  It has the effect of completely skipping the jump instruction.  
Consider the following sequence of x86 assembly instructions 
\texttt{\{BB1:cmpl \$0,\%eax; je BB3; BB2:jmp BB4; BB3:...; BB4:...\}} as shown in Fig.~\ref{fig:x86_BB}. In a fault-free environment, the CPU is expected to execute \texttt{BB1;BB3} when \texttt{\%eax} is equal to 0, and execute \texttt{BB1;BB2} otherwise. 
However, in the presence of hardware faults, either (or both) of the jump instructions (\texttt{je BB3} and \texttt{jmp BB4}) may be turned into a \emph{nop}, thus leading to new (and abnormal) execution paths.

We call it \emph{branch instruction skipping}.
It is a transient error instead of a permanent error.  For example, if the jump instruction is inside a loop, it is possible to skip the jump instruction during one iteration but not during another iteration. 
Furthermore, branch instruction skipping is possible not only for the unconditional jump instruction (\texttt{jmp}), but also for conditional jump instructions such as \texttt{je} and \texttt{jne}. In all these cases, the jump instruction is interpreted as a null operation (\textit{nop}).

\section{Our Method}
\label{sec:methodology}

We now present the top-level procedure of our method in this section and highlight how it differs from the baseline symbolic execution algorithm.  Detailed algorithms of the subroutines will be presented in the next two sections. 

\subsection{The Notations}

Before presenting the top-level procedure, we define the relevant notations.
We consider a victim program $P$ with a set $L$ of program locations, where locations $l_{\mathit{init}}$ and $l_{\mathit{end}}$ are the start and end of the program, respectively.  Furthermore, $l_{\mathit{bad}}\in L$ represents the error location.  In other words, reaching $l_{\mathit{bad}}$ is a safety violation. This way of representing a safety violation is without loss of generality, since in a Turing-complete programming language, a safety violation can always be expressed as an embedded assertion. Furthermore, \texttt{assert(c)} may be modeled as \texttt{if(!c)\{$l_{\mathit{bad}}$\}}.

In the victim program's control flow graph (CFG), the nodes represent locations in $L$, and the edges represent events.  Each event is a tuple $t=(l, inst, l')$ where $l,l'\in L$ and $inst$ is an instruction. 
An execution path is a sequence of events $t_1,t_2,\dots,t_n$.  
If $t_n.l' = l_{\mathit{bad}}$ (the last location is an error location), it means that the execution path ends with a safety violation. 
In contrast,  $t_n.l' = l_{\mathit{end}}$ means that the execution path ends without a safety violation.

The instruction $inst$ may be one of two types: it is either an \emph{assignment} or a \emph{branching statement}. This assumption is also made without loss of generality because any event encountered during an execution may be modeled by instructions of these two types.
Given a set $V$ of program variables, an assignment is denoted $v:=expr$, where $v\in V$ is the left-hand-side (lhs) variable and $expr$ is the right-hand-side (rhs) expression, defined over the variables in $V$.  
A branching statement is denoted $assume(c)$, where $c$ is a conditional expression, also defined over the variables in $V$. 
For example, \texttt{if(x>5)\{\}else\{\}} in a C program corresponds to two branching statements in our formalism, denoted by $assume(x>5)$ and $assume(x\leq 5)$, respectively.

\newcommand{\Ind}[1]{\hspace{#1ex}\hspace{#1ex}\hspace{#1ex}}
\newcommand{\textcode}{\texttt}
\newcommand{\true}{true}
\newcommand{\pcon}{pcon}
\newcommand{\mem}{mem}

\begin{algorithm}[t]
    \caption{Symbolic execution with (and without) our redundancy pruning techniques.}
    \label{alg:baseline}
 {\footnotesize
    \begin{algorithmic}[1]
    \STATE \textbf{Initialize:} Run \textsc{Explore}($s_0$) with the initial state stack $S \leftarrow \{s_0\}$
    \STATE \textsc{Explore}( state $s$ ) \{ 
\textcolor{purple}{ 
    \STATE {\Ind{1}     {\bf if} ( \textsc{PruningCondition}($s$) is satisfied)  \textcolor{gray}{\ \ //use fault count and path summary to skip paths}  }
    \STATE {\Ind{2}      {\bf return;}                                }
}
    \STATE {\Ind{1}    $S$.push($s$);                                                             }
    \STATE {\Ind{1}    {\bf if } ($s$ is a branching point)                                       }
    \STATE {\Ind{2}      {\bf foreach } ($t \in s.branch$)                             }
    \STATE {\Ind{3}         $s' \leftarrow$ \textsc{NextState}($s,t$)                            }
    \STATE {\Ind{3}         \textsc{Explore}($s'$);                                               }
    \STATE {\Ind{1}    {\bf else if} ($s$ is an internal node)                                    }
    \STATE {\Ind{2}      $s' \leftarrow$ \textsc{NextState}($s,t$)                               }
    \STATE {\Ind{2}      \textsc{Explore}($s'$);                                                  }
    \STATE {\Ind{1}    {\bf else} //end of an execution path                                      }
    \STATE {\Ind{2}      compute a program input;                                        }
\textcolor{purple}{
    \STATE {\Ind{2}      \textsc{UpdateSuffixSummary}($S$);     \textcolor{gray}{\ \ \ \ \ \ \ \ \ \ \ //summarize the path (suffix)}    }
}
    \STATE {\Ind{1}    $S$.pop();                                                                 }
   \STATE \}
    \STATE {\textsc{NextState}( state $s$, event $t$ ) \{                                                         }
    \STATE {\Ind{1} $\langle l, \pcon, \mem \rangle  \leftarrow$ the state $s$           }
    \STATE {\Ind{1} $\langle l, inst, l' \rangle \leftarrow$ the event $t$                 }
    \STATE {\Ind{1} {\bf if} ( $inst$ is $\mathbf{assume(c)}$ )                                      }
\textcolor{purple}{
    \STATE {\Ind{2}      \textsc{UpdateFaultCounter}($s,t$);    \textcolor{gray}{\ \ \ \ \ \ \ \ \ \ //compute fault count}    }
}
    \STATE {\Ind{2}      $s' \leftarrow \langle l', \pcon\wedge c, mem  \rangle$                 }
    \STATE {\Ind{1} {\bf else if} ( $inst$ is \textbf{assignment} $v:=expr$ )                    }
    \STATE {\Ind{2}      $s' \leftarrow \langle l', \pcon, \mem[v \mapsto expr] \rangle$         }
    \STATE {\Ind{1} {\bf return} $s'$; }
    \STATE \}
    \end{algorithmic}
 }
\end{algorithm}

\subsection{Baseline Symbolic Execution}
\label{subsec:baseline}

Our new symbolic execution algorithm is implemented as an extension of KLEE~\cite{KLEE08}, which is a widely used symbolic execution engine for C/C++ programs built on the intermediate representation (IR) of the popular LLVM compiler~\cite{LLVM04}. 
Specifically, if we ignore the highlighted lines (3-4, 15, and 22) in Algorithm~\ref{alg:baseline}, then it describes the \emph{baseline} symbolic execution procedure as implemented in KLEE. 
Thus, in the remainder of this section, we closely follow the description used by KLEE and other symbolic execution tools~\cite{CUTE05,SAGE12}. 

As shown in Line~1 of Algorithm~\ref{alg:baseline}, starting from the initial state $s_0$ of the program $P$, the recursive function \textsc{Explore}$(s_0)$ performs a depth-first-search (DFS) of the feasible program paths. 
At each step, \textsc{Explore}$(s)$ picks an instruction $inst$ that can be executed at the current state $s$, computes the next state $s'$ (Lines~8 and~11), and invokes \textsc{Explore} on $s'$ recursively (Lines~9 and~12).

Let the execution path be $\pi = s_0, s_1, \ldots, s_n$.  During symbolic execution, $\pi$ is stored in the state stack $S$, which in turn is leveraged to explore all execution paths systematically. 
After the procedure finishes executing an execution path (Lines~13--15), it computes a program input. This is useful because, if the path ends with a safety violation, the program input will be able to reproduce the violation.

The state is a tuple $s=(l, pcon, mem)$ where $pcon$ is the path condition that must be satisfied for the execution to reach location $l\in L$, and $mem$ is the symbolic memory map. That is, $mem[v]$ stores, for each variable $v\in V$, its corresponding value expression. 
Given the current state $s$, and more importantly, its program location $s.l$, we know all the information needed to compute the next state $s'$.  This includes which events are enabled at $s$.  These events are represented by the outgoing edges of the node $s.l$ in the program's control flow graph.

Finally, in the subroutine \textsc{NextState}$(s)$, there are two cases. For an assignment $v:=expr$ (Line~24), the new path condition $pcon$ will remain the same, while the symbolic memory map will be updated such that $mem[v]$ holds $expr$. 
For a branching statement $assume(c)$ (Line~23), the new path condition will be updated to $pcon\wedge c$ while the new symbolic memory map ($mem$) will remain the same.

\subsection{Our New Symbolic Execution Procedure}

If we include the highlighted lines in Algorithm~\ref{alg:baseline}, then Algorithm~\ref{alg:baseline} describes our new symbolic execution algorithm, which leverages two important techniques for mitigating the path explosion problem.  Specifically, Lines~15 and~22 compute the information needed for identifying redundant executions based on the current execution path $\pi$, while Lines~3--4 utilize this information to avoid the redundant executions.

\subsubsection{Pruning Based on Fault Saturation}

Our first redundancy removal technique relies on the fact that the number of activated faults in any execution path must be less than or equal to the fault budget $\beta$. Thus, if at any moment during symbolic execution, the currently executed path (stored in the state stack $S$) has already activated more than $\beta$ faults, symbolic execution is terminated.
Before applying this technique, however, we must perform a program transformation, denoted $P' \leftarrow \textsc{FaultModeling}(P,\beta)$ where the newly transformed program $P'$ is symbolically executed instead of the original program $P$. 

Details of the program transformation will be presented in Section~\ref{sec:modeling}. Here, it suffices to say that, for each branching statement $br\in Br$ in the original program $P$, we add a new branch $br'$ in the transformed program $P'$, such that $br'$ is controlled by a \emph{fault flag} denoted \texttt{bFT}, whose value is either $true$ or $false$.  We also maintain a \emph{fault counter} denoted \texttt{FC}, whose value is increased every time a fault flag such as \texttt{bFT} is activated (set to $true$). 

Together, these auxiliary variables (\texttt{bFT}, \texttt{FC} and $\beta$) model the aforementioned requirements that (1) the original branch $br$ is skipped (turned to \textit{nop}) if and only if its corresponding fault flag \texttt{bFT} is set to $true$ and (2) an execution path is explored if and only if  $\mathtt{FC} \leq \beta$ holds, meaning that the fault count does not exceed the fault budget.

\subsubsection{Pruning Based on Path Summary}

Our second redundancy removal technique relies on the fact that an execution path $\pi:=\pi_{pre}\pi_{post}$ is the concatenation of a prefix $\pi_{pre}$ and a suffix $\pi_{post}$. 
Given another path $\pi':=\pi'_{pre}\pi'_{post}$, it is possible for $\pi$ and $\pi'$ to have different prefixes ($\pi_{pre}\neq \pi'_{pre}$) but a common suffix ($\pi_{post}=\pi'_{post}$).  
In such a case, if the common suffix $\pi_{post}$ has been symbolically executed following $\pi_{pre}$, does it need to be symbolically executed again following $\pi'_{pre}$?  The answer depends on whether symbolically executing $\pi_{post}$ again can lead to \emph{previously-unexplored} program behavior.

For example, a sufficient condition for answering \emph{no} to the above question is when $(s=s')$.  Here, $s$ and $s'$ are the two states reached by the two prefixes $\pi_{pre}$ and $\pi'_{pre}$, respectively.  
This technique is known as state hashing/merging, but it has limited success in practice, for two reasons. 
First, since $s$ and $s'$ are symbolic expressions, checking if they are semantically equivalent is expensive, meaning that the overhead may not be compensated by the time saved. More importantly, it is rare for the two states ($s$ and $s'$) to be exactly the same.

In contrast, we use a sufficient condition that is significantly different and more general than state hashing/merging.  As shown in Lines~15 and~3 of Algorithm~\ref{alg:baseline}, the technique consists of two parts.
First, we compute (Line~15) a succinct path summary for the \emph{explored} suffix starting from each program location $l$, stored in a map entry $\mathtt{WP}[l]$.
Then, we check if the pruning condition $\pcon \rightarrow \mathtt{WP}[l]$ holds (Line~3).  If the pruning condition holds, it means $pcon \subseteq \mathtt{WP}[l]$.  Since the path condition for the prefix $\pi'_{pre}$, denoted $pcon$, is fully covered by the path summary $\mathtt{WP}[l]$, we can safely avoid executing the common suffix again.

\vspace{1ex}
In the next two sections, we present the detailed algorithms of our new techniques for fault modeling and path pruning, respectively. 
\section{Accurate Fault Modeling via Program Transformation}
\label{sec:modeling}

We propose a program transformation technique for accurately modeling the impact of fault attacks on the victim program.  Given the original program $P$, we construct a new program $P'$ such that $P'$ is semantically equivalent to $P$ in a fault-free execution environment, but has all of the additional behaviors under fault attacks. 

\begin{figure}
\centering
    \centering
    \includegraphics[width=\linewidth]{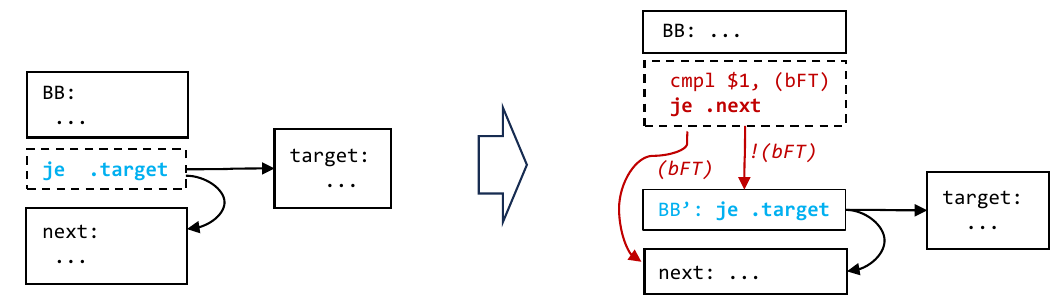}
    \caption{Illustrating how Lines 6-7 of Algorithm~\ref{alg:modelfaults} work.
Given a basic block $\mathtt{BB}$ shown on the left-hand side, our program transformation algorithm automatically adds the new control flow edge \textcolor{purple}{$br'$:= \textbf{if}(\texttt{bFT}) \{\textbf{goto}} \texttt{next}\textcolor{purple}{\}} shown on the right-hand side, right before \textcolor{cyan}{$br:= \texttt{je .target}$}.}
    \label{fig:fault_modeling}
\end{figure}

\subsection{The Fault Modeling Algorithm}

Algorithm~\ref{alg:modelfaults} shows our fault modeling subroutine, which takes a program $P$ as input and returns a new program $P'$ as output. 
Initially, $P'$ is set to be the same as $P$. 
Then, for each branching instruction $br$ in $P$, we model the impact of \emph{instruction skipping}  in two steps. 
First, we add a \emph{fault flag} $\mathtt{bFT}$ to the new program $P'$. 
Second, we create a new branching instruction $br'$ controlled by $\mathtt{bFT}$, with $\mathtt{next}$ as the target basic block. That is,  $br' :=$ \textbf{if}($\mathtt{bFT}$) \{\textbf{goto} $\mathtt{next}$\}.
Finally, we add $br'$ to the new program $P'$, at the program location right before the original $br$, inside the basic block $\mathtt{BB}$ that holds $br$.

\begin{algorithm}
    \caption{ Subroutine $P' \leftarrow$ \textsc{FaultModeling} $(P)$ for our new program transformation.}
    \label{alg:modelfaults}
 {\footnotesize
    \begin{algorithmic}[1]
        \STATE \textbf{Initialize:} new program $P' \leftarrow$ a copy of program $P$
        \FOR{\textbf{each} ( branching instruction \textcolor{cyan}{$br \in Br$} in program $P$ )}
            \STATE Add a fault flag, denoted $\mathtt{bFT}$,  to program $P'$
            \STATE $\mathtt{BB} \leftarrow$  the basic block that holds \textcolor{cyan}{$br$}
            \STATE $\mathtt{next} \leftarrow$ the basic block that immediately follows \textcolor{cyan}{$br$} in the program code
            \STATE Create \textcolor{purple}{$br' :=$ \textbf{if}($\mathtt{bFT}$) \{\textbf{goto}} $\mathtt{next}$\textcolor{purple}{\}}
            \STATE Add \textcolor{purple}{$br'$} to program $P'$, right before \textcolor{cyan}{$br$} inside $\mathtt{BB}$
        \ENDFOR
        \STATE \Return $P'$;
    \end{algorithmic}
 }
\end{algorithm}

Fig.~\ref{fig:fault_modeling} illustrates how Algorithm~\ref{alg:modelfaults} works, where the basic block of $P$ containing the branch $br$ is shown on the left-hand side.  The corresponding basic blocks of the newly transformed program $P'$ are shown on the right-hand side, where the added instructions and edges are highlighted in red.
%
Without loss of generality, in this figure, we only illustrate how the conditional jump instruction is handled.  The unconditional jump instruction can be regarded as a special case of conditional jump where the transition from $\mathtt{BB}$ to $\mathtt{next}$ is always disabled.

As illustrated in Fig.~\ref{fig:fault_modeling}, regardless of whether $br$ is conditional (e.g., \texttt{je target}) or unconditional (e.g., \texttt{jmp target}), it must have $\mathtt{BB}$, $\mathtt{target}$ and $\mathtt{next}$. 
Here, $\mathtt{BB}$ is the basic block to which $br$ belongs,
$\mathtt{target}$ is the basic block where $br$ jumps to, and 
$\mathtt{next}$ is the basic block immediately following $br$ in the program code. 
%
For a concrete example of the $\mathtt{next}$ basic block, please refer to \texttt{BB3} in the middle subfigure of Fig.~\ref{fig:motivating}: \texttt{BB3} is the basic block immediately following \texttt{BB2} in the program code.

\subsection{Applied to the Running Example}

\begin{figure*}[t]
\centering
    \includegraphics[width=\linewidth]{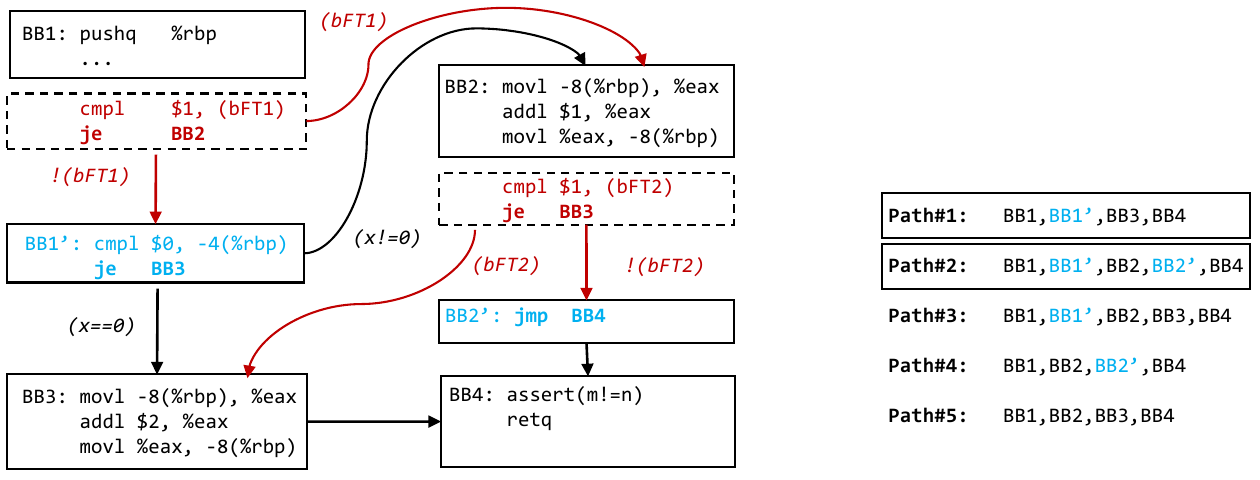}
    \caption{Illustrating the advantages of our new method on the running example: on the left-hand side is the transformed program $P'$ for the original program $P$ in Fig.~\ref{fig:motivating}, where fault-induced control flows are highlighted as red instructions and edges. The five paths explored by baseline symbolic execution (without pruning) are shown on the right-hand side; only the first two paths will be fully explored by our new algorithm (with pruning).}
    \label{fig:motivating_transformed}
\end{figure*}

We now use another example, i.e., the program in Fig.~\ref{fig:motivating},  to explain our fault modeling algorithm in detail.  Recall that in Fig.~\ref{fig:motivating}, the right-hand side shows the control flow graph of the original program $P$. 
Accordingly, in Fig.~\ref{fig:motivating_transformed}, the left-hand side shows the control flow graph  of the new program $P'$.

Specifically, for the conditional branch \texttt{je BB3} colored in \emph{blue} in Fig.~\ref{fig:motivating_transformed}, we add the new branch \texttt{je BB2}, which is controlled by the fault flag \texttt{bFT1}.  When \texttt{bFT1}$=true$, meaning that a fault is injected, the control transfers to \texttt{BB2} without checking the original condition \texttt{(x!=0)}. 
Similarly, for the unconditional branch \texttt{jmp BB4}, we add the new branch \texttt{je BB3}, which is controlled by the fault flag \texttt{bFT2}. When \texttt{bFT2}$=true$, meaning that another fault is injected, the control transfers to \texttt{BB3} instead of the original destination \texttt{BB4}.

Overall, the new program $P'$ shown on the left-hand side of Fig.~\ref{fig:motivating_transformed} has five paths.  They are listed on the right-hand side of this figure.  In contrast, the original program $P$ has only two paths; 
they are the first and the second of the five paths
listed in Fig.~\ref{fig:motivating_transformed}. 

\vspace{1ex}
\noindent
\textbf{Advantage over the Prior Work.}
To summarize, our fault modeling technique is significantly more accurate than the current state-of-the-art~\cite{ESOP23}, which models the impact of a fault by inverting the test flag.  For the running example, in particular, the existing technique would include only 4 of the 5 paths, but would miss the  path \texttt{BB1$\rightarrow$BB2$\rightarrow$BB3$\rightarrow$BB4}.  Arguably, this is the most interesting path, since it visits both the then-branch and the else-branch of the \textsf{if-else} statement during one execution of the program. 
While it may be hard to imagine for a novice developer, as we have mentioned earlier, it may occur in practice due to \emph{branch instruction skipping}. 
The current state-of-the-art technique will miss this violation, as well as many other violations that can be detected by our new method (see the experimental results in Section~\ref{sec:experiment}). 

\vspace{1ex}
%
We now state the soundness and completeness of our technique. 
\begin{theorem}
\label{thm:modeling}
Our fault modeling technique as presented in Algorithm~\ref{alg:modelfaults} is both sound and complete in that the transformed program (1) includes all real violations and (2) does not include bogus violations.
\end{theorem}
\begin{proof}
The soundness and completeness of our fault modeling technique follow directly from the program transformation steps presented in Algorithm~\ref{alg:modelfaults}. As we have illustrated in Fig.~\ref{fig:fault_modeling}, the algorithm guarantees that all existing control flow paths of the program are retained in the transformed program.
Furthermore, in the transformed program, all of the newly added control flow paths correspond to faults that lead to skipping the branch instructions.
Therefore, the transformed program includes all real violations and no bogus violations.
\end{proof}

\section{Efficient Symbolic Execution for Fault Analysis}
\label{sec:pruning}

We now present the two pruning techniques for speeding up fault analysis based on symbolic execution. As mentioned earlier, one technique is based on bounding the maximal number of activated faults during an execution, and the other technique is based on skipping previously-explored suffixes. 
In Algorithm~\ref{alg:baseline}, these two techniques correspond to the subroutines \textsc{UpdateFaultCounter}, \textsc{UpdateSuffixSummary}, and \textsc{PruningCondition}.

\subsection{Pruning Based on the Fault Count}
\label{subsec:leveraging1}

If we ignore Lines 8--9 in Algorithm~\ref{alg:pruning}, then Algorithm~\ref{alg:pruning} implements the pruning technique based on bounding the maximum number of activated faults during an execution.
Specifically, the subroutine \textsc{UpdateFaultCounter} updates the fault counter during the symbolic execution of a program path, while the subroutine \textsc{PruningCondition} checks the fault bounding condition $(s.\mathtt{FC}>\beta)$ to decide if symbolic execution can terminate early for the program path.

\begin{algorithm}
    \caption{Subroutines that support efficient fault bounding and redundancy pruning.}
    \label{alg:pruning}
 {\footnotesize
    \begin{algorithmic}[1]
    \STATE {\textsc{UpdateFaultCounter}( state $s$, event $t$ ) \{                             }
    \STATE {\Ind{1}    {\bf if} ( instruction $t.inst$ is of the type \textbf{assume}($\mathtt{bFT}=true$) ) }
    \STATE {\Ind{2}      $s.\mathtt{FC} \leftarrow s.\mathtt{FC}+1$ } 
    \STATE \}
    \STATE \textsc{PruningCondition}( state $s$ ) \{ 
\textcolor{purple}{
    \STATE {\Ind{1}    {\bf if} ( $s.\mathtt{FC} >$ the fault budget $\beta$ ) \textcolor{gray}{\ \ \ \ \ //pruning based on the fault count $s.\mathtt{FC}$}}
    \STATE {\Ind{2}      {\bf return $true$;}                                }
    \STATE {\Ind{1}    {\bf else if} ( $s.\pcon \wedge \neg \mathtt{WP}[s.l]$ is unsatisfiable ) \textcolor{gray}{\ \ \ \ \ //pruning based on the path summary $\mathtt{WP}[s.l]$}}
    \STATE {\Ind{2}      {\bf return $true$;}                        }
}
    \STATE {\Ind{1}    {\bf else } }
    \STATE {\Ind{2}      {\bf return $false$;}                        }
    \STATE \}
    \end{algorithmic}
 }
\end{algorithm}

Recall that the state stack $S$ stores the current execution path as a sequence $s_0\xrightarrow{t_1}s_1\xrightarrow{t_2}\ldots \xrightarrow{t} s$. 
At the initial state $s_0$, the value of the fault counter $s_0.\mathtt{FC}$ is set to 0.
Inside \textsc{UpdateFaultCounter}($s,t$),  every time a fault is activated, by executing the special branch statement \textbf{assume}($\mathtt{bFT}=true$), the value of the fault counter $s.\mathtt{FC}$ increases by 1. 
Thus, $s.\mathtt{FC}$ is the total number of activated faults along the current path. 

For the reason stated above, inside \textsc{PruningCondition}$(s)$, we can end the symbolic execution of the current path when $s.\mathtt{FC}> \beta$ (Lines~6--7), meaning that the total number of activated faults exceeds the fault budget.
Thus, pruning is guaranteed to be sound (or safe), meaning it never skips a path that should not be skipped.

Another advantage of our method over existing methods is that, when there are multiple ways of injecting faults to violate a safety property, our method tries to return the one with fewer faults as follows.
On top of the baseline symbolic execution algorithm (Section~\ref{subsec:baseline}), which aims at exploring the program paths in a strictly depth-first search (DFS) order, at every branching point, our method symbolically explores the fault-free path (where $\mathtt{bFT}=false$) before symbolically exploring the faulty path (where $\mathtt{bFT}=true$).
For each fault-controlled branching instruction, it means that our method always explores the inactive branch \textbf{assume}(\texttt{!bFT}) before it explores the active branch \textbf{assume}(\texttt{bFT}).

\subsection{Pruning Based on the Path Summary} 
\label{subsec:leveraging2}

If we include Lines 8--9 in Algorithm~\ref{alg:pruning}, then the pruning condition leverages both the fault counter and the path summary.
Specifically, the path condition $s.pcon$ captures the set of states that can be reached by executing the new prefix (from $s_0$ to $s$). 
\blue{At the same time, the suffix summary $\mathtt{WP}[s.l]$  captures the weakest precondition computed from \emph{all the explored suffixes starting from $s$}.}
Thus, if  $s.pcon\wedge\neg \mathtt{WP}[s.l]$ is unsatisfiable, it means that the set of states captured by $s.pcon$ is a subset of the set of states captured by $\mathtt{WP}[s.l]$. Consequently, continuing the execution from $s$ cannot lead to any previously unexplored behavior.

\begin{figure*}[t]
        \centering
        \includegraphics[width=1\textwidth]{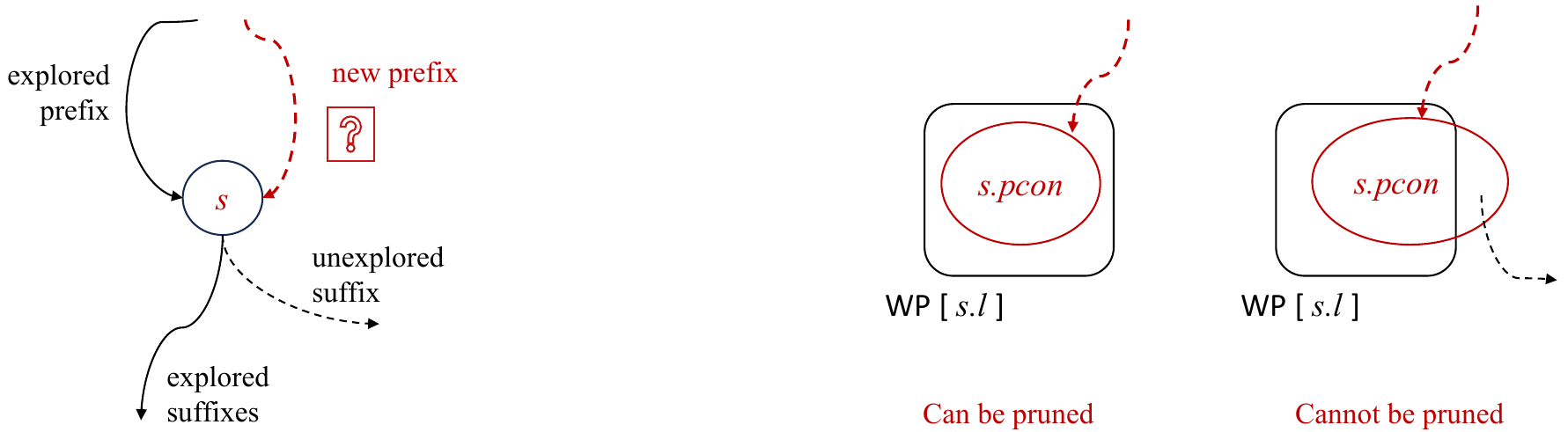}
        \caption{Illustrating how our pruning algorithm works: to decide if we can safely prune the current execution (new prefix) shown on the left, we check the pruning condition $s.pcon\rightarrow \mathtt{WP}[s.l]$, which may lead to one of the two cases shown on the right.}
        \label{fig:WP_pruning}
\end{figure*}

Fig.~\ref{fig:WP_pruning} illustrates the situation on the left-hand side.  Depending on whether $s.pcon$ is fully included in $\mathtt{WP}[s.l]$, there are two cases, as shown on the right-hand side of Fig.~\ref{fig:WP_pruning}.  
While both $s.pcon$ and $\mathtt{WP}[s.l]$ are symbolic expressions (or logical formulas represented in the format of SMT constraints), they capture two sets of program states.  
That is, program state belongs to $s.pcon$ if and only if that state makes $s.pcon$ evaluate to $true$.

We use Z3 to check if $s.pcon\wedge\neg\mathtt{WP}[s.l]$ is unsatisfiable. Being unsatisfiable means that the set of program states captured by $s.pcon$ is indeed a subset of the set of program states captured by $\mathtt{WP}[s.l]$. 
\blue{In other words, if we continue the current execution from state $s$, we will only explore the known behaviors.}
Since our goal is to avoid exploring the previously-explored behaviors, we can safely end the current execution early.  
This is the reason why pruning is guaranteed to be sound, meaning that it never skips executions that should not be skipped.

\subsection{Updating the Path Summary}

We now present the subroutine \textsc{UpdateSuffixSummary}. Internally, the subroutine has two steps: (1) computing the weakest precondition ($wp$) along each explored path, and (2) combining them to form a summary ($\mathtt{WP}$) of all explored paths.

\subsubsection{Weakest Precondition ($wp$)}
Following Dijkstra~\cite{dijkstra1976wp}, we compute the \emph{weakest precondition} of an execution path by traversing the path backwardly.
Let the path be $\pi=s_0\xrightarrow{t_1}s_1\xrightarrow{t_2}\ldots \xrightarrow{t_n}s_n$.  The last program location $s_n.l$ is either $l_{end}$ (meaning the path ends without violation) or $l_{bad}$ (meaning the path ends with a violation).
Results of the weakest precondition computation are stored in a map named $wp$, where each entry $wp[s_n.l]$ is a symbolic expression for the program location $s_n.l$ associated with the state $s_n$. 
\begin{itemize}
\item 
If $s_n$ is the last state in $\pi$, then set $wp[s_n.l]=true$. 
\item 
For any other state $s_i$ where $0\leq i<n$,  if the corresponding instruction $t_i.inst$ is of the type \textbf{assume}($c$), then set $wp[s_i.l] = wp[s_{i+1}.l]\wedge c$. 
\item 
For any other state $s_i$ where $0\leq i<n$,  if the corresponding instruction $t_i.inst$ is of the type $v:=expr$, then set $wp[s_i.l] = wp[s_{i+1}.l](v\mapsto expr)$.
\end{itemize} 
Here, $(v\mapsto expr)$ means that, inside $wp[s_{i+1}.l]$, we replace the left-hand-side (lhs) variable $v$ with the right-hand-side (rhs) expression $expr$.

\subsubsection{From $wp$ to the Union ($\mathtt{WP}$)}
%
%
Given a set of execution paths, we first compute $wp$ for each path, and then union them together.   The result, denoted $\mathtt{WP}$, is a summary of all paths. 
Note that the union is performed at each program location $s_i.l\in L$, not at each state $s_i\in S$. 
Thus, the resulting $\mathtt{WP}$ is a map where $\mathtt{WP}[s_i.l]$ stores the union of all the individual $wp[s_i.l]$ expressions. 
%

Specifically, after computing $wp_1, \cdots, wp_k$ for $k$ paths, we combine them to create the union $\mathtt{WP}$ by applying  
{
    \[
        \mathtt{WP}[l] \equiv \bigvee_{1\leq j \leq k} wp_j[l]
    \]
}%
for every program location $l\in L$ in the program.

\subsubsection{Applied to the Running Example}
We now illustrate how $wp$ is computed along a path, and how $wp$'s are combined to compute $\mathtt{WP}$ using the example program in Fig.~\ref{fig:motivating}. 
For ease of presentation, we assume that the fault budget $\beta$ is set to 2. 
Under this assumption, there are five paths in the CFG of the transformed program, shown on the right-hand side of Fig.~\ref{fig:motivating_transformed}.

To compute $wp_1$ for Path \#1, for example, we first initialize $wp_1$ to $true$ at the end of the path.  Then, we traverse the path backward. After encountering \textbf{assume}($c$), we apply the rule $wp[s_i.l] = wp[s_{i+1}.l]\wedge c$ to obtain 
{
\[
    wp_1[\texttt{assert}] \equiv true \wedge (m \neq n).
\]
}%
Next, we encounter \texttt{n+=2} in \texttt{BB3}. By applying the rule $wp[s_i.l] = wp[s_{i+1}.l](v\mapsto expr)$, we obtain
{
\[
    wp_1[\texttt{n+=2}] \equiv wp_1[\texttt{assert}](n\mapsto n+2) \equiv (m \neq n+2).
\]
}%
Next, we encounter \texttt{BB1':je BB3} (i.e., \textbf{assume}$(x=0)$), for which we obtain
{
\[
    \begin{aligned}
        wp_1[\texttt{BB1':je BB3}] &\equiv wp_1[\texttt{n+=2}] \wedge (x=0)  \equiv (m \neq n + 2) \wedge (x=0).
    \end{aligned}
\]
}%
Finally, we obtain the entire $wp_1$ map as follows:
{
\[
    \begin{aligned}
        wp_1[\texttt{assert}] &\equiv (m \neq n), \\
        wp_1[\texttt{n+=2}] &\equiv (m \neq n+2), \\
        wp_1[\texttt{BB1':je BB3}] &\equiv (m \neq n + 2) \wedge (x=0), \\
        wp_1[\texttt{je BB2}] &\equiv (m \neq n + 2) \wedge (x=0) \wedge \neg bFT1.
    \end{aligned}
\]
}%
Since there is only one explored path for the moment, we set $\mathtt{WP} = wp_1$.



Next, the symbolic execution executes Path \#2, starting at where it is forked at \texttt{je BB3}.
When executing the assertion statement in \texttt{BB4} again at the end of Path \#2, we apply the existing $WP[assert]$ to check against the current path condition.
At this point, we have 
\[
s.pcon \equiv (m = 3) \wedge (n = 1) \wedge \neg bFT1 \wedge (x\neq 0) \wedge \neg bFT2,
\]
and 
\[
\mathtt{WP}[assert] \equiv (m \neq n).
\]
We further check if the new path condition,
\[
\begin{aligned}
    s.pcon \land \neg \mathtt{WP},
\end{aligned}
\]
which simplifies to
\[
(\neg bFT1 \land \neg bFT2 \land x\neq 0) \land \neg (3 \neq 1).
\]
Since the above constraint equals $false$,  we stop executing Path \#2 at the assertion statement and mark it as pruned.
We then repeat the same process to compute the path summary for Path \#2, starting with
\[
wp_2[\texttt{assert}] \equiv true \wedge (m \neq n).
\]

Eventually, we have $\mathtt{WP}[l] \equiv wp_1[l] \vee \ldots \vee wp_5[l]$, and Paths 2, 4, and 5 are partially executed, due to the path summary based redundancy pruning in our method.

\vspace{1ex}
%
We now state the soundness of our technique. 
\begin{theorem}
\label{thm:pruning}
Our path pruning technique presented in this section is sound. That is, paths that are pruned away by our technique are guaranteed to be redundant.
\end{theorem}
\begin{proof}
The soundness of our path pruning technique can be established as follows.
First, recall that our path pruning technique relies on two methods: fault saturation and weakest precondition. 
For fault saturation, by definition, our pruning technique would only eliminate paths where the number of activated faults exceeds a predefined threshold. 
For weakest precondition (WP), whether a path should be pruned away is determined by checking if the WP-based path summary is covered by the current symbolic program state: if the answer is yes, it means that all possible extensions of the current path prefix have been covered by the summary of previously explored paths.
Therefore, in both cases, paths that are pruned away by our technique are guaranteed to be redundant.
\end{proof}


\section{Experiments}
\label{sec:experiment}

We have implemented our method as a software tool by leveraging the LLVM compiler~\cite{LLVM04}, the KLEE symbolic execution engine~\cite{KLEE08}, and the Z3 SMT solver~\cite{Z308}.
After compiling a C program to LLVM bit-code, the tool first conducts fault modeling and then applies symbolic execution. 
Our fault modeling technique was implemented as an LLVM optimization pass, which transforms the original program $P$ to the new program $P'$ at the LLVM IR level. 
Our redundancy pruning technique was implemented as an extension of KLEE, where we first compute fault count and WP-based path summary, and then use Z3 to check the pruning conditions.

To allow a fair comparison with the state-of-the-art method, we have re-implemented the fault modeling and pruning techniques of \cite{ESOP23} using KLEE.
To allow an ablation study, we have also implemented a baseline of our symbolic execution method, which includes our fault modeling and fault bounding techniques, but excludes our summary-based pruning technique.
%
%
In total, our implementation adds 4,000 lines of C++ code to LLVM and KLEE.

\subsection{Benchmark Programs}
\label{subsec:benchmarks}

We have evaluated our tool on 112 benchmark C programs.
They include all of the 8 versions of VerifyPIN~\cite{ESOP23}, which is a widely-used benchmark suite for fault attack analysis~\cite{muArchiFI23,FISCC16,verifyPIN22,faultPLDI24,faultPOPL24}.
In addition, they also include three \textit{complete} sets of benchmark programs from SV-COMP 2025's \emph{ReachSafety} category, namely \emph{bitvector}, \emph{bitvector-loops}, and \emph{array-crafted}.
%
These three sets of benchmarks were designed to challenge verification tools on handling low-level bit-accurate computations (such as integer overflow and bitwise operations) crucial to embedded systems, instead of the less accurate, standard mathematical integer verification for general-purpose programs. 
Thus, they align well with our research objectives of determining if hardware fault attacks cause abnormal control flows in embedded software, leading to the reachability of an error state. 

Note that nine programs from the \textit{bitvector} category (\textit{jain\_*.c}) were excluded because they contain infinite loops without branching or assertions, and thus are outside our research scope.
Moreover, since KLEE is designed for test case generation, instead of proving that an assertion always holds, its capability of handling unbounded loops is limited.
%
%
%
%
Therefore, we set a limit on the maximal number of symbolic execution steps for KLEE. 
Our goal is to ensure that KLEE terminates reasonably quickly, so we can have a baseline for comparing our pruning method with the current state-of-the-art method.

The statistics of the 112 benchmark programs are shown in Table~\ref{tab:benchmark-categories}.
Column 1 shows the category name and Column 2 shows the total number of programs in the category.
Columns 3--5 show the minimum, maximum, and average number of lines of code (LoC) for all programs in a category.
Column 6 shows a brief description of the benchmark programs.
Note that the LoC span a wide range, indicating that the programs have varying complexity; they include small programs with $<$50 LoC as well as larger programs with $>$700 LoC.

\begin{table}[t!]
    \centering
    \caption{Statistics of the control-intensive benchmark C programs used in the evaluation.}
    \label{tab:benchmark-categories}
    \resizebox{\textwidth}{!}{
        \begin{tabular}{|l|c|c|c|c|p{0.38\linewidth}|}
        \hline
        \multirow{2}{*}{\textbf{Category Name}} & \multirow{2}{*}{\textbf{Num.\ Programs}} & \multirow{2}{*}{\textbf{Min. LoC}} & \multirow{2}{*}{\textbf{Max. LoC}} & \multirow{2}{*}{\textbf{Avg. LoC}} & \multirow{2}{*}{\textbf{Benchmark Description}}    \\
         & & & & & \\
 \hline\hline
        VerifyPIN                               & 8           & 80               & 141              & 100.9             & a set of benchmark programs widely used to evaluate fault attacks      \\\hline
        array-crafted                           & 43          & 44               & 86               & 54.5              & SV-COMP benchmarks: programs with integer and array operations                     \\
        bitvector                               & 20          & 24               & 143              & 80.5              & SV-COMP benchmarks: programs with bounded integer operations
\\
        bitvector (unbounded)                     & 41         & 66              & 733              & 524.5               & SV-COMP benchmarks: programs with bounded integers and unbounded loops       \\ \hline
        \end{tabular}
    }
\end{table}

\subsection{Evaluation Method}

The experiments were designed to answer the following questions:
\begin{itemize}
\item 
\blue{Is our fault modeling technique more effective than the current state-of-the-art, thus allowing us to detect more real violations and fewer bogus violations?}
\item
\blue{Is our new pruning technique effective in reducing the overall computational cost of the symbolic execution based fault analysis?} 
\end{itemize}

To evaluate the quality of our fault modeling technique, we compared it with the existing method on all benchmark programs.
We set the maximum execution depth to 50 when exploring a single symbolic path for the benchmarks in \textit{bitvector (unbounded)} to bound the execution of unbounded loops. 
The only exceptions are for programs \textit{soft\_float\_3.c.cil.c} and \textit{soft\_float\_3a.c.cil.c}, where we lowered the maximum depth to 30. This is because the two programs are more complex than others: the baseline method in KLEE cannot finish analyzing them when the execution depth is too large.

Our experiments were conducted on a computer with a 4.7 GHz AMD R9 7900X CPU and 32 GB of RAM running the Ubuntu 22.04 LTS Linux system. 
We set a 90-minute timeout per program.
In the remainder of this section, we present the experimental results obtained to answer these questions.

\subsection{Results on Fault Modeling}
\label{subsec:rq1}

We first present a summary of the safety violations found  by different methods in Table~\ref{tab:effect-summary},
where Column 1 shows the category name and Column 2 shows the total number of programs in the category.  
Columns 3-4 show the results of the existing method, including the number of programs where a violation was not found, and the number of programs where the violation was found. 
Similarly, Columns 5-6 show the results of our method.
The last row shows the total number of programs, the ones without violations, and the ones with violations. 
%
Overall, our method found 14 more programs with violations than the existing method (90 violations versus 76 violations); furthermore, a careful analysis shows that all of the 14 violations are real violations.

For example, our method found a violation in \emph{bAnd1.c} while the existing method found no violation.  A closer look at the program input computed by KLEE showed that it is a realistic violation. 
Conversely, our method found no violation in \emph{gcd\_2.c} while the existing method found a violation. However, a closer look at the program input computed by KLEE showed that it is a bogus violation.
In both cases, the differences support the claim that our method more accurately models the impact of fault attacks on a victim program.

\begin{table}[t]
    \centering
    \caption{Results on fault modeling: the total number of programs for which safety violations were found by the existing method~\cite{ESOP23}, our baseline method, and our optimized method, respectively. \# No Violation is the number of programs on which a violation was not found, and \# Found Violation is the number of programs on which a violation was found.}
    \label{tab:effect-summary}

    \scalebox{0.75}{
        \begin{tabular}{|l|c|cc|cc|}
        \hline
        \multirow{2}{*}{\textbf{Category Name}} & \multirow{2}{*}{\textbf{\# Programs}} & \multicolumn{2}{c|}{\textbf{Existing Method~\cite{ESOP23}}} & \multicolumn{2}{c|}{\textbf{Our Method}}                                   \\ \cline{3-6} 
        & & \multicolumn{1}{c|}{\textbf{\# No Violation} } & \textbf{\# Found Violation} & \multicolumn{1}{c|}{\textbf{\# No Violation} } & \textbf{\# Found Violation}  \\ \hline\hline

        VerifyPIN                    & 8     & \multicolumn{1}{c|}{3}                             & 5                    & \multicolumn{1}{c|}{3}                             & 5                    \\\hline

        array-crafted                & 43    & \multicolumn{1}{c|}{2}                             & 41                    & \multicolumn{1}{c|}{1}                             & 42                  \\
        bitvector                    & 20    & \multicolumn{1}{c|}{3}                             & 17                    & \multicolumn{1}{c|}{1}                             & 19                  \\
        bitvector (unbound)            & 41    & \multicolumn{1}{c|}{28}                            & 13                    & \multicolumn{1}{c|}{17}                            & 24                  \\ \hline
        \textbf{Total}               & 112   & \multicolumn{1}{c|}{36}                            & \textbf{76}           & \multicolumn{1}{c|}{22}                            & \textbf{90}         \\ \hline
        \end{tabular}
    }
\end{table}

We now present detailed results for the VerifyPIN programs in Table~\ref{tab:effect-verifypin}, where Columns 1--2 show the name of each program and the number of injected faults, Column~3 shows the result of the existing method, and Column~4 shows the result of our method. For each method, we indicate whether a violation is detected ({\vNO} means no and \vYES\ means yes). For these eight programs, our method detected the same number of violations as the existing method. That is, with at most one injected fault, a violation was found in the first five of the eight programs.  When the number of injected faults was increased to 2, both our method and the existing method detected violations in all programs, including the last three of the eight programs.

\begin{table}[t]
    \centering
    \caption{\blue{Detailed results for VerifyPIN benchmarks, where the two methods found the same number of violations.  Here,  {\vNO } means violation was not found, and {\vYES } means violation was found. }}
    \label{tab:effect-verifypin}

    \resizebox{\textwidth}{!}{
        \begin{tabular}{|l|c|c|c|}
        \hline
        \textbf{Program Name\ \ \ \ \ \ \ \ }  & \textbf{\# Faults} & \textbf{Violation Found by Existing Method} & \textbf{Violation Found by Our Method}  \\ \hline
        VerifyPIN\_0.c & 1               & \vYES                                  & \vYES \\
        VerifyPIN\_1.c & 1               & \vYES                                  & \vYES \\
        VerifyPIN\_2.c & 1               & \vYES                                  & \vYES \\
        VerifyPIN\_3.c & 1               & \vYES                                  & \vYES \\
        VerifyPIN\_4.c & 1               & \vYES                                  & \vYES \\
        VerifyPIN\_5.c & 1               & \vNO                                   & \vNO  \\
        VerifyPIN\_6.c & 1               & \vNO                                   & \vNO  \\
        VerifyPIN\_7.c & 1               & \vNO                                   & \vNO  \\ \hline
        VerifyPIN\_5.c & 2               & \vYES                                  & \vYES \\
        VerifyPIN\_6.c & 2               & \vYES                                  & \vYES \\
        VerifyPIN\_7.c & 2               & \vYES                                  & \vYES \\ \hline
        \end{tabular}
    }
\end{table}

We present detailed results for a subset of SV-COMP benchmarks in Table~\ref{tab:effect-all-diff}.  It only includes programs where a violation was found by one method but not by the other method.
Furthermore, {\vYES } means the found violation was real, whereas (\vvYES) means the found violation was bogus (in the sense that it would not have occurred in practice).
Overall, our method found a violation in 15 of the 16 programs, whereas the existing method found a violation in only 1 of the 16 programs.

\begin{table}[t]
    \centering
    \caption{Detailed results for the subset of SV-COMP benchmarks where a violation was found by one method but not by the other. Here, {\vNO } means no violation was found, and {\vYES } means a violation was found. Furthermore, (\vvYES) means the violation was bogus (would not occur in practice).}
    \label{tab:effect-all-diff}
    \resizebox{\textwidth}{!}{
        \begin{tabular}{|l|c|c|c|c|}
        \hline
        \textbf{Program Name}             & \textbf{\# Faults} & \textbf{Violation Found by Existing Method} & \textbf{Violation Found by Our Method}   \\ \hline
        bAnd1.c                   & 1               & \vNO                                     & \vYES  \\
        gcd\_2.c                  & 1               & (\vvYES)                                     & \vNO   \\
        gcd\_3.c                  & 1               & \vNO                                     & \vYES  \\
        interleave\_bits.c        & 1               & \vNO                                     & \vYES  \\
        s3\_clnt\_1.BV.c.cil-1a.c & 1               & \vNO                                     & \vYES  \\
        s3\_clnt\_2.BV.c.cil-1.c  & 1               & \vNO                                     & \vYES  \\
        s3\_clnt\_2.BV.c.cil-1a.c & 1               & \vNO                                     & \vYES  \\
        s3\_clnt\_3.BV.c.cil-1.c  & 1               & \vNO                                     & \vYES  \\
        s3\_clnt\_3.BV.c.cil-1a.c & 1               & \vNO                                     & \vYES  \\
        s3\_clnt\_3.BV.c.cil-2.c  & 1               & \vNO                                     & \vYES  \\
        s3\_clnt\_3.BV.c.cil-2a.c & 1               & \vNO                                     & \vYES  \\
        soft\_float\_2.c.cil.c    & 1               & \vNO                                     & \vYES  \\
        soft\_float\_2a.c.cil.c   & 1               & \vNO                                     & \vYES  \\
        soft\_float\_5.c.cil.c    & 1               & \vNO                                     & \vYES  \\
        sum02-1.c                 & 1               & \vNO                                     & \vYES \\
        bor2.c                    & 2               & \vNO                                     & \vYES \\ \hline
        \end{tabular}
    }
\end{table}

Furthermore, a closer look (using the code snippet in Fig.~\ref{fig:bogus-violation}) shows that the violation found by the existing method was bogus, whereas the violations found by our method were all real.
The bogus violation was due to \emph{test inversion}, an inaccurate fault modeling technique. 

If inside the function \texttt{gcd\_test()} the condition \texttt{(b!=0)} were inverted, the while-loop would be skipped entirely. Furthermore, if this occurred when  \texttt{a=12} and \texttt{b=8}, the resulting \texttt{a=12} would violate \texttt{assert(b>a)}.
However, 
this bogus violation cannot occur in the real world.
With our more accurate fault modeling, the while-loop would never be skipped entirely. 
After executing the loop body at least once, the assertion would never be violated.

%
\begin{figure}
\begin{lstlisting}[
    language=C,basicstyle=\footnotesize\ttfamily,keywordstyle=\color{blue},
    xleftmargin=.02\textwidth, backgroundcolor=\color{white}
]
signed char gcd_test (signed char a, signed char b) {
    signed char t;
    if (a < (signed char)0) 
        a = -a;
    if (b < (signed char)0) 
        b = -b;
    while (b != (signed char)0) {
        t = b; 
        b = a % b; 
        a = t;
    }
    return a;
}
\end{lstlisting}
\caption{
The code snippet (in gcd\_2.c) shows that, while a bogus violation is reported due to test inversion (an inaccurate fault modeling technique), the bogus violation is avoided by using our new fault modeling technique.
}
\label{fig:bogus-violation}
\end{figure}

This is because, after the while-loop is compiled to machine code, the loop body is guarded by a conditional jump to the loop exit, e.g., \texttt{jz .exit} for \texttt{while(b!=0)}. When the loop condition \texttt{(b!=0)} is not satisfied, the jump will be taken to skip the loop body; but when the loop condition is satisfied, the jump will not be taken and the execution will fall through to the loop body.
When a hardware fault turns the jump instruction into \textit{nop}, the execution will also fall through to the loop body.  Therefore, an attacker cannot physically force the while-loop to be skipped when the loop condition \texttt{(b!=0)} is satisfied. 

To summarize, these experimental results show that our new fault modeling technique is significantly more effective than the existing method.

\subsection{Results on Redundancy Pruning}

\begin{figure}[t]
\centering
\begin{minipage}{.495\textwidth}
\centering
\includegraphics[width=\textwidth]{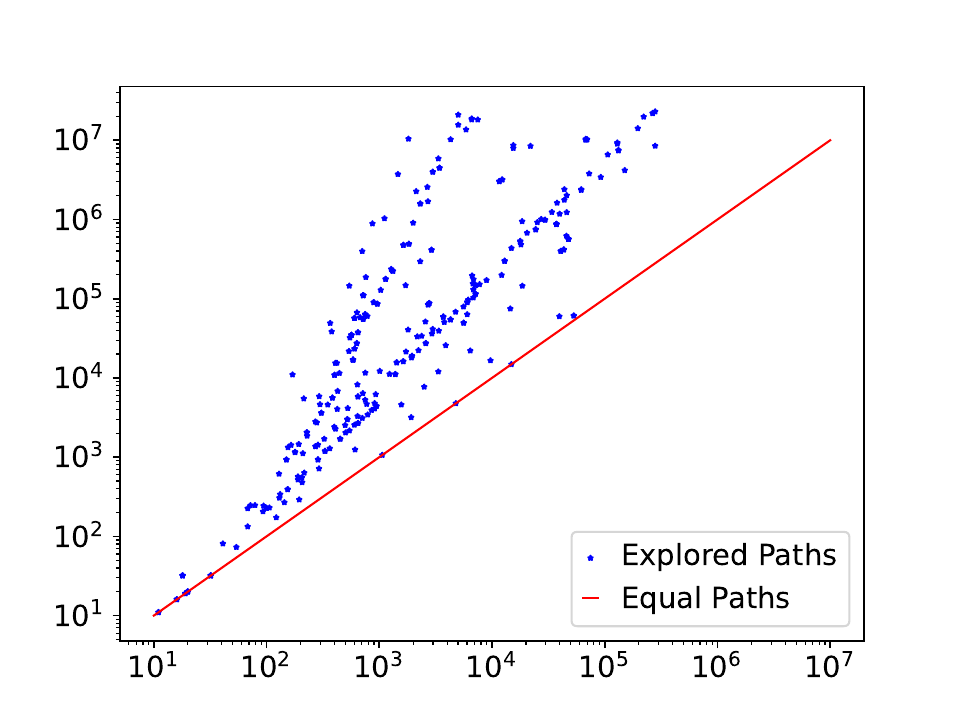}
\subcaption{Number of explored paths:  with our pruning \\ 
technique ($x$-axis) versus the baseline  ($y$-axis)}
\label{fig:explored_paths}
\end{minipage}
\begin{minipage}{.495\textwidth}
\centering
\includegraphics[width=\textwidth]{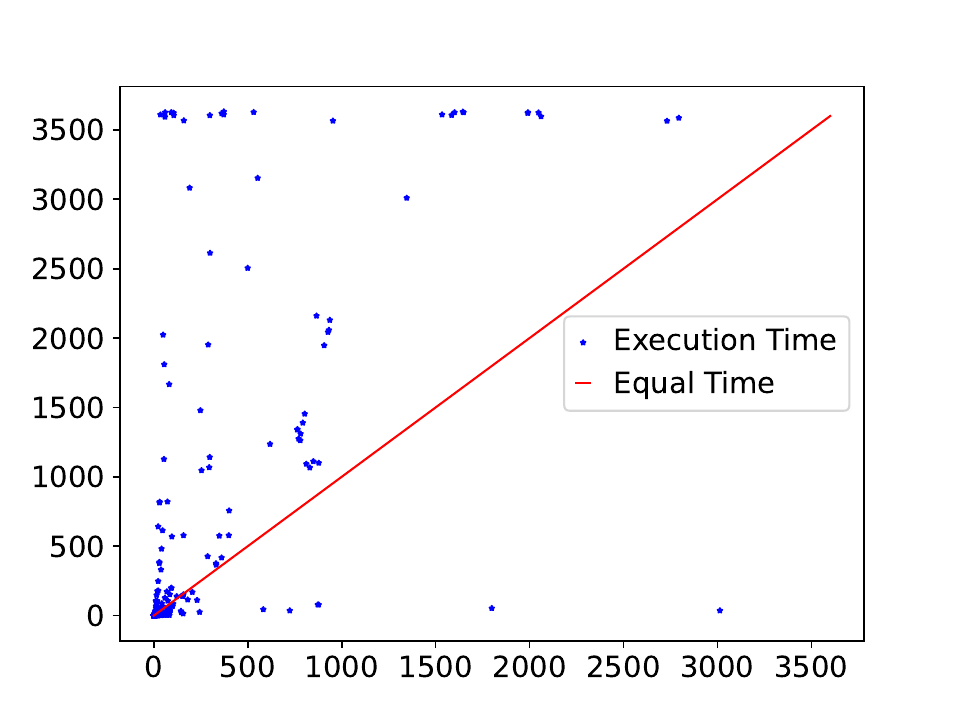}
\subcaption{Execution time in seconds: with our pruning \\
technique ($x$-axis) versus the baseline ($y$-axis)}
\label{fig:execution_time}
\end{minipage}

\caption{Experimental results that demonstrate the advantage of our pruning technique over baseline symbolic execution: the number of explored paths (left) and the execution time (right) of our symbolic execution method, with and without the new pruning technique. In both scatter plots, blue points above the red diagonal lines are winning cases for our new pruning technique.}
\label{fig:efficiency}
\end{figure}

%

To evaluate the effectiveness of our redundancy pruning technique, we compared two aspects of the experimental results between our baseline and optimized methods: the number of explored paths and the execution time.
To obtain more data, we increased the maximum fault budget from 1 to 4 for benchmarks in the categories \emph{VerifyPIN}, \emph{array-crafted}, and \emph{bitvector}.
During this process, we stopped increasing the number of injected faults whenever timeouts occurred.
The reason why we excluded \emph{bitvector (unbounded)} from this step is explained below.

First, the optimized method already outperformed the baseline method in terms of execution time across nearly all cases in \emph{bitvector (unbounded)} when the fault budget was set to one.
Therefore, increasing the number of faults would only yield limited additional insights.
Second, the benchmarks in \emph{bitvector (unbounded)} are more complex, and using a higher fault budget would require additional computational resources and time, which may not be feasible.

We present the number of explored paths and the running time in the scatter plots in Fig.~\ref{fig:efficiency}.
Since the number of explored paths falls in a wide range, we present the results on a log scale.
Here, the number of explored paths is shown in Fig.~\ref{fig:explored_paths}, where the baseline is on the $y$-axis and our optimized method is on the $x$-axis;
%
the running time (in seconds) is shown in Fig.~\ref{fig:execution_time}, where the baseline is on the $y$-axis and our optimized method is on the $x$-axis. 
In both sub-figures, blue points above the red diagonal lines are winning cases for our optimized method.

The results in Fig.~\ref{fig:explored_paths} show that our optimized method almost always explores fewer paths than the baseline. Furthermore, in many cases, it explores significantly fewer paths. 
The results in Fig.~\ref{fig:execution_time} show that our optimized method almost always takes less time, although there are exceptions that show up as dots below the diagonal line. This is due to the overhead of WP-based pruning.  Nevertheless, they typically occur when the running time is short; when the running time is long, our optimized method wins.

There are some cases in which our optimized method takes notably longer than the baseline, particularly for the benchmarks \emph{gcd\_2.c} and \emph{gcd\_3.c} when the number of faults was set to 3 and 4; the corresponding data points are located in the bottom-right corner of Fig.~\ref{fig:execution_time}.
We inspected these benchmark programs and found that most of the time was spent on invoking solvers to determine the satisfiability of WP-based pruning constraints.
But, overall, the results in Fig.~\ref{fig:execution_time} show that our pruning technique is effective in reducing the computational cost.



\section{Related Work}
\label{sec:related}


\blue{Our method relies on symbolic execution to accurately and efficiently analyze the impact of hardware faults on software.}
\blue{While in theory, any software program may become the victim of hardware fault attacks, in practice, the most popular targets are embedded software programs~\cite{Bar-ElCNTW06,YuceSW18} and cryptographic software programs~\cite{AESInstSkip12,LFACryptAttack23}.}  These programs are the targets of our method.
As mentioned earlier in this paper, \blue{\binsec}~\cite{ESOP23} is the most closely related prior work, but it is not as effective as our method in terms of fault modeling,  as demonstrated by our experimental evaluation.
%

\blue{Besides {\binsec}~\cite{ESOP23}, which represents the current state-of-the-art, there are various earlier works on fault analysis~\cite{PattabiramanNKI13,LeHGD18,Given-WilsonJL20,LacombeFBP24,Lancia22,LarssonH07}. However, they leverage existing techniques like static analysis and concolic execution in a more or less straightforward fashion; in particular, they do not focus on improving the underlying symbolic analysis procedures using redundancy pruning techniques.}

Although symbolic execution is a popular technique for analyzing both sequential and concurrent software~\cite{SymExeWP15,GuoWW17,AdversarialSymbolic18} and it has been implemented in many existing tools including CUTE~\cite{CUTE05}, KLEE~\cite{KLEE08} and SAGE~\cite{SAGE12}, it is known to suffer from the path explosion problem. 
Various techniques have been developed to mitigate path explosion, including forward analysis techniques such as state merging~\cite{KuznetsovKBC12} and backward analysis techniques such as post-conditioned pruning~\cite{PostSymbolicJournal,PostSymbolic,GuoKW16}.
However, none of these existing techniques specifically target software programs under fault attacks.

The reason why we focus on symbolic execution is because our goal in this work is to detect safety violations in software programs, for which symbolic execution is particularly strong. If the goal were to generate proofs of no safety violations, other techniques that focus primarily on verification (instead of falsification) could be used instead.  Specifically, such techniques can be used to prove fault-resistance~\cite{MoroHER14,muArchiFI23,SongFuFaultCrypto23,PesinBMP25}.  For example, Tollec et al.~\cite{muArchiFI23} applied bounded model checking to software programs running on a RISC-V processor to analyze the impact of faults injected at the micro-architectural level, while  Tan et al.~\cite{SongFuFaultCrypto23} used SAT solvers to evaluate the fault-resistance of cryptographic circuits.  Our method is different in that it is geared toward detecting violations and, more importantly, it is a software-only solution where the impact of hardware faults is abstracted into the \emph{branch instruction skipping} threat model.

\blue{Beyond software-only solutions and tools, there is a body of work on physically conducting fault injection using power~\cite{PowerGlitching05} and clock glitching~\cite{ClockGlitch18}, laser beams~\cite{LaserBeam03} on hardware boards, as well as studying the impact of the injected faults on software programs using hardware-software co-simulation~\cite{GrycelS21,LiuSS24}.}
These empirical studies have led to the \emph{branch instruction skipping} threat model used in our method~\cite{Bar-ElCNTW06,YuceSW18} as well as other threat models including \emph{memory bit flipping}~\cite{BitFlip17,BitFlip20,Rowhammer19}. While we focus exclusively on modeling instruction skipping in this work, we foresee no technical difficulty in extending our method to the \emph{memory bit flipping} threat model.


\section{Conclusion}
\label{sec:conclusion}


We have presented a symbolic execution based method to analyze the impact of hardware faults on the safety of a software program both accurately and efficiently. 
The method has two new techniques. The first one is a compiler-based program transformation technique for accurate fault modeling.  The second one is a redundancy pruning technique that leverages fault saturation and weakest precondition to avoid symbolically executing redundant program paths.
\blue{Our experimental evaluation on two sets of benchmark programs shows that the method significantly outperforms the current state-of-the-art method, in that it can detect previously missed violations and can significantly reduce the overall computational cost.}




\bibliographystyle{plainurl}
\bibliography{src/main}

\end{document}